\documentclass[twocolumn,showpacs,prd,floatfix]{revtex4-1}
\pdfoutput=1

\usepackage{graphicx,xcolor}
\usepackage{amssymb,amsmath}

\newcommand{\beq}{\begin{equation}}
\newcommand{\eeq}{\end{equation}}
\newcommand{\bea}{\begin{eqnarray}}
\newcommand{\eea}{\end{eqnarray}}
\newcommand{\R}{\mathbb{R}}

\newcommand{\n}{{(n)}}
\newcommand{\na}{{(1)}}
\newcommand{\nb}{{(2)}}

\newcommand{\barg}{\bar g}
\newcommand{\barA}{\bar A}

\begin{document}

\title{
Solving the Hamiltonian constraint for 1+log trumpets
}

\author{Tim Dietrich, Bernd Br\"ugmann}

\affiliation{Theoretical Physics Institute, 
University of Jena, 07743 Jena, Germany}

\date{September 12, 2013}

\begin{abstract}
The puncture method specifies black hole data on a hypersurface with
the aid of a conformal rescaling of the metric that exhibits a
coordinate singularity at the puncture point. When constructing
puncture initial data by solving the Hamiltonian constraint for the
conformal factor, the coordinate singularity requires special
attention. The standard way to treat the pole singularity occurring in wormhole
puncture data is not generally applicable to trumpet puncture data. We
investigate a new approach based on inverse powers of the
conformal factor and present numerical examples for single punctures of the wormhole and
1+log-trumpet type. Additionally, we describe a method to solve the Hamiltonian
constraint for two 1+log trumpets for a given extrinsic curvature with non-vanishing
trace. We investigate properties of this constructed initial data during binary black
hole evolutions and find that the initial gauge dynamics is reduced. 
\end{abstract}


\maketitle

\section{Introduction}
\label{introduction2009}

A central issue in numerical general relativity is how to handle black
holes and their spacetime singularities. In general, we can choose
different foliations which avoid the singularities, either by explicit
singularity excision \cite{SeiSue92} or by the puncture
method \cite{BraBru97,CamLouMar05,BakCenCho05}, which leads to
wormhole or trumpet slices that avoid the singularity
automatically. The moving puncture method of
\cite{CamLouMar05,BakCenCho05} combines black hole punctures
\cite{BraBru97,Bru97} with the Baumgarte-Shapiro-Shibata-Nakamura (BSSN)
formalism \cite{ShiNak95,BauSha98b} and appropriate gauge choices for
the lapse \cite{BonMasSei94a} and the shift
\cite{AlcBruPol01,AlcBruDie02}. This leads to robust 
black hole simulations for a large variety of black hole systems.
Most initial data for such configurations are created with the help of
maximally sliced wormholes, e.g.\ \cite{BraBru97,AnsBruTic04,BroLow04}.
When evolved with 1+log slicing by the moving puncture method, the
wormholes lose contact to their second asymptotically flat end and are
deformed into a trumpet (loosely speaking, ``half'' a wormhole)
\cite{HanHusPol06,Bro07a,HanHusOMu06,HanHusOhm08}.  The
quasi-equilibrium state of a moving puncture is a trumpet.  It is
remarkable how quickly and robustly the gauge handles the transition
from maximally sliced wormholes to 1+log sliced trumpets.

In this paper we address the question of how to compute trumpet
initial data. It is a curious fact that we can easily obtain trumpets
from wormholes by evolution, but there is no method yet to pose
initial data representing two moving and spinning black holes by
trumpets directly, without evolution, on appropriate slices in the
1+log gauge. 
Our analytic understanding of trumpets is restricted essentially to
spherical symmetry, i.e.\ the Schwarzschild trumpet
\cite{HanHusPol06}, for which both 1+log and maximal slicing is known
\cite{BauNac07,HanHusOMu06,HanHusOhm08}. 
Although puncture evolutions employ the 1+log gauge, so far most
investigations of trumpet initial data beyond Schwarzschild have
focussed on maximally sliced trumpets 
\cite{ImmBau09,HanHusOMu09,Bau12a}. 
Constant mean curvature slices with trumpets were considered in \cite{BucPfeBar09}.
In \cite{BauEtiLiu08} it is shown that constructing 1+log trumpet data
for orbiting black holes may fail if one 
assumes the existence of a helical Killing vector, since in general such data
is not asymptotically flat. However, this does not rule out the existence
of asymptotically flat 1+log trumpet data that is only approximately stationary,
which is the case we are interested in.

The goal in this paper is to analyse and resolve some of the 
difficulties that arise when solving the Hamiltonian constraint of
the 3+1 initial data problem for 1+log trumpets. We postpone the treatment
of the momentum constraint. Let us summarize the key issues.
Consider the conformal transverse-traceless (CTT)
decomposition \cite{Yor79}, where the physical metric is obtained by a conformal
rescaling of a given background metric, $g_{ij} = \psi^4 \barg_{ij}, $
where $\psi>0$ is the conformal factor. 
For puncture data we can assume that the metric is conformally flat.
The basic feature is that the conformal
factor has a coordinate singularity at $r=0$, where $r$ denotes the
Cartesian distance to the puncture. Initial data for Schwarzschild can
be written as wormhole puncture data with two asymptotic infinities, or
as trumpet puncture data 
extending from a sphere with minimal area inside the horizon to
infinity. The coordinate singularity for $r\rightarrow0$ takes the form 
\beq
  \psi_{wormhole} \sim \frac{1}{r}, 
  \quad \psi_{trumpet} \sim \frac{1}{\sqrt{r}}.  
\label{coordsing}
\eeq 
The above generalizes to more than one puncture by including a pole
for each puncture.
The Hamiltonian constraint for conformally flat initial data
can be written as $\Delta \psi + F \psi^5 + G\psi^{-7} = 0,$
with the boundary condition that $\psi \rightarrow 1$ as
$r\rightarrow\infty$. Here $\Delta$ is the flat-space Laplace
operator, and the functions $F$ and $G$ are obtained from the extrinsic curvature.
Given the singular behavior of the conformal factor,
(\ref{coordsing}), the question is how to compute such irregular
solutions of the Hamiltonian constraint when a numerical solution is required. 

For wormhole data with Bowen-York (BY) extrinsic curvature \cite{BowYor80}, a
successful strategy \cite{BraBru97} is to write the conformal factor as
$  \psi = \psi_S + u$ assuming
$\Delta\psi_S = 0$ on $\R^3\setminus \{0\}$,
where $\psi_S$ is the singular, but analytically known solution for
vanishing extrinsic curvature for one or more punctures (the
Brill-Lindquist conformal factor). We obtain a solution 
since $\Delta(1/r)=0$ for $r\neq0$.
However, for trumpet data we encounter
$\Delta(1/\sqrt{r})\neq0$. Furthermore, for 1+log trumpet data the
trace of the extrinsic curvature, $K$, does not vanish. This
introduces additional issues compared to maximal slicing, where $K=0$.

In this paper we explore an alternative to $\psi = \psi_S + u$.
Rather than attempting a split into singular and regular pieces, we
consider a power of the inverse $1/\psi$ of the conformal
factor~\cite{Gun10a,Bau12a}. For example, if $\psi\sim1/\sqrt{r}$,
then $\psi^{-4}\sim r^2$ is a regular function at the puncture.
Introducing
\beq
  \psi = f^p, \quad p < 0,
\label{psiq}
\eeq
for $p = -1/2$ we have $f_{wormhole}\sim r^2$, and for
$p=-1/4$ we have $f_{wormhole}\sim r^4$ and $f_{trumpet}\sim r^2$,
cf.\ (\ref{coordsing}).

The question is whether anything has been gained when writing the
Hamiltonian constraint in terms of the new conformal factor $f$.
Although taking the inverse of $\psi$ to some
power raises the differentiability at the puncture, a priori it is not
clear whether the singularity has only been shifted to other terms.
Indeed, 
\beq
  \Delta f^p = p f^{p-1} \left(\Delta f + (p-1) \frac{(\nabla f)^2}{f}\right).
\label{Deltafp}
\eeq
The question is whether there are numerical issues for $f$ approaching
zero at the puncture, e.g.\ whether the numerical derivative
in $\frac{(\nabla f)^2}{f}$ vanishes sufficiently fast for a regular
result.
Here $p=1$, which corresponds to the power $4$ in the conformal
transformation of the metric, 
is precisely the choice
that avoids first order derivatives in the Hamiltonian constraint.
We will discuss how the numerical quality of the initial data
depends on $p$.

Inverse powers of $\psi$ have appeared in different contexts, for
example in black hole evolutions \cite{CamLouMar05}. 
We suggested their use in the
Hamiltonian constraint for the thesis of Gundermann \cite{Gun10a}. That
work focused on 1+log trumpets and on uniqueness issues 
of related model problems. For example, in a simple case
($F=const.$) the solutions are not unique and 
two solutions were constructed explicitly. Some numerical
experiments were performed in \cite{Gun10a} as well, although a robust
numerical implementation was missing. This is one of the goals of this paper.

Baumgarte \cite{Bau12a} suggested the same approach with inverse
powers of the conformal factor, as well as a working numerical scheme
based on 3d finite differences. The present paper and \cite{Bau12a}
are complementary in that we work with a 3d pseudospectral method that
is a variant of \cite{AnsBruTic04}.
A suitable numerical implementation that can handle possible
regularity issues for the new conformal factor is found in both
numerical approaches.

However, the major difference between the present work and other
numerical studies of trumpet initial data \cite{Bau12a,HanHusOMu09,Bau11,AnsBai13} 
is that we do not assume maximal slicing 
(nor do we attempt to impose helical symmetry \cite{BauEtiLiu08}). 
Instead of maximal trumpets with $K=0$ and $F=0$, we consider 1+log
trumpets with $K\neq0$ and $F\neq0$, which adds the $\psi^5$ term in
the Hamiltonian constraint.

The paper is organized as follows. In Sec.\ \ref{HamCon} we introduce
the Hamiltonian constraint, discuss how the wormhole puncture method works
and what the problems are extending this method to trumpets.
In Sec.\ \ref{RewHam} we rewrite the Hamiltonian constraint 
with a regular conformal factor and describe some numerical calculations
for a single wormhole, for a single 1+log trumpet, and for multiple 
1+log trumpets. In Sec.\ \ref{Evo} we compare evolutions starting with
the standard maximal wormhole data, with maximal trumpet data, and with
the new 1+log trumpet data. We conclude in Sec.\ \ref{Con}.

\section{Hamiltonian constraint for punctures}
\label{HamCon}

\subsection{Hamiltonian constraint}

The standard 3+1 decomposition \cite{Yor79} is formulated in terms of a
three-metric $g_{ij}$ and its extrinsic curvature $K_{ij}$. The
Hamiltonian constraint for vacuum is
\beq
  R(g) + K^2 - K_{ij}K^{ij} = 0,
\eeq
where $R(g)$ is the Ricci scalar of $g_{ij}$, and $K=g^{ij}K_{ij}$ is the
trace of the extrinsic curvature. The conformal
transverse-traceless (CTT) decomposition introduces a conformal factor
$\psi>0$ such that the ``physical'' metric $g_{ij}$ is obtained from
the ``conformal'' metric $\barg_{ij}$ by
\beq
  g_{ij} = \psi^4 \barg_{ij}. \label{conformaltransformation}
\eeq
We could insert (\ref{psiq}) here, $g_{ij}=f^{4p}\barg_{ij}$, but we may as well
insert the new conformal rescaling in the final decomposition so that we
do not have to repeat the standard calculations.
Since 
\beq
  R(\psi^4 \barg) = \psi^{-4}R(\barg) - 8 \psi^{-5} \bar\Delta \psi,
\eeq
the Hamiltonian constraint for the conformal variables becomes
\beq
   \bar\Delta \psi - \frac{1}{8} R(\barg_{ij}) \psi 
   + \frac{1}{8} (K_{ij}K^{ij} - K^2) \psi^5 = 0, 
\label{ham0}
\eeq
where the Laplace operator $\bar\Delta$ refers to the conformal metric.
When constructing initial data, $\barg_{ij}$ is considered given while
$\psi$ is determined as a solution of (\ref{ham0}). 
For puncture data we assume that the conformal metric is flat, which
implies $R(\barg_{ij}) = 0$. 
We drop the overhead bar to simplify our notation when it is evident
that we are referring to the conformal variables.

The CTT decomposition also introduces the
tracefree part of the extrinsic curvature, $A_{ij}=K_{ij}-g_{ij}K/3$,
and the conformal transformation
\beq
  A_{ij} = \psi^{-2} \barA_{ij}, \quad K = \bar K.
\eeq
In this work we do not address the question of how to
solve the momentum constraint. Instead we consider examples where the
extrinsic curvature is given, either as an explicit solution to the
momentum constraint, or as an approximation that does not solve the momentum
constraint, but that provides an initial guess for its
solution at a later stage.

In the following we consider Bowen-York extrinsic curvature, for which
$K=0$, as well as 1+log trumpet data, for which $K\neq0$. The Hamiltonian
constraint takes the form
\beq 
\Delta \psi + F \psi^5 + G\psi^{-7} = 0.
\label{ham1}
\eeq
For Bowen-York extrinsic curvature, $F=0$ and
$G=\barA_{ij}\barA^{ij}/8$. For trumpet data, we can set
$F=(K_{ij}K^{ij} - K^2)/8$ and $G=0$ without performing the
trace decomposition, or $F=-K^2/12$ and $G=\barA_{ij}\barA^{ij}/8$. 
Furthermore, one could also add Bowen-York extrinsic curvature
to trumpet data in order to imbue the trumpet with linear and angular
momentum. 

\subsection{Wormhole puncture}
\label{WormPunc}

The original puncture method is motivated by the Schwarzschild metric in
spatially isotropic coordinates on slices of constant Schwarzschild time. 
The metric is conformally flat with vanishing extrinsic curvature,
\beq
\psi=\psi_S = 1 + \frac{m}{2r}, \quad g_{ij} = \psi_S^4 \delta_{ij}, \quad
K_{ij}=0.
\label{Spuncture}
\eeq
The conformal factor is the fundamental solution of the (flat-space) Laplace
equation,
\beq
\Delta \psi = 0,
\label{ellSpuncture}
\eeq
on a ``punctured'' $\R^3$, i.e.\ $\R^3$ without the puncture point
$r=0$, subject to the boundary condition 
$\psi(\infty)\equiv\lim_{r\rightarrow\infty}\psi=1$. As a function on
$\R^3$, $\psi$ has a coordinate singularity at $r=0$. In terms of the
Schwarzschild radial coordinate $R$, however, we have the geometric picture of a
wormhole between two asymptotically flat regions which is isometric
about the horizon at $r=m/2$, $R=2m$. As $r$ ranges from 0 to
$\infty$, $R$ drops from $\infty$ at the inner asymptotically flat
end to a minimum at the throat for $R=2m$ before growing again to
$\infty$ at the outer asymptotically flat end.

If we consider a puncture with Bowen-York extrinsic curvature, 
then we have to solve 
\beq
  \Delta \psi + G \psi^{-7} = 0,
\label{ellBYpuncture}
\eeq
where $G = O(r^{-6})$ if there is spin, and $G=O(r^{-4})$ if there is
linear momentum. While analytically not an issue (e.g.\ existence and
uniqueness can be proven), the question is how to treat the coordinate
singularity at the puncture numerically.

The original puncture method of \cite{BraBru97,BeiOMu94,BeiOMu96,DaiFri01} 
proceeds by considering
\bea
\psi &=& \psi_S + u, \label{puncturetrick0} \\
\Delta u + G (\psi_S + u)^{-7} &=& 0, \label{puncturetrick1} \\
u(\infty) &=& 0. 
\eea
The key observation is that there exists a unique regular solution $u$
on (the un-punctured) $\R^3$. Put simply, we can solve for a
regular function $u$ on the entire $\R^3$, add it to the singular
background solution $\psi_S$ for vanishing extrinsic curvature, and
obtain a solution with spin and momentum. Since the pole of $\psi_S$
has been handled analytically by using $\Delta\psi_S=0$ in the
transition from (\ref{ellBYpuncture}) to (\ref{puncturetrick1}), we do not
expect and in practice do not encounter numerical difficulties when
solving (\ref{puncturetrick1}) for $u$.

Although the ``puncture trick''
(\ref{puncturetrick0}-\ref{puncturetrick1})
is straightforward as presented, some of its features
should be recalled since they are relevant to the construction of
trumpet data. 
Obviously the puncture method depends on the existence of an analytic
solution for the Schwarzschild solution. By using $\psi_S$ or its
immediate generalization to multiple Brill-Lindquist punctures, we
enforce the existence of black holes in the data. The treatment of the
puncture point is somewhat subtle. First we discover solutions to
(\ref{ellSpuncture}) or (\ref{ellBYpuncture}) that have a pole at
$r=0$, introduce $u=\psi-\psi_S$ on $\R^3\setminus\{0\}$, realize that
$u$ is uniquely determined on $\R^3$ by (\ref{puncturetrick1}) (where
we have compactified the inner infinity), and declare this to be the
unique solution we want. In technical terms, there is a removable
singularity at $r=0$. However, by choosing the unique extension we
also make a choice about the inner boundary. From the point of view of
the wormhole construction, we could be working on $\R\times S^2$ with
$\psi=1$ at both asymptotic ends, but this would not automatically
build in a black hole of mass $m$.

Other features of the puncture solution depend on the choice of
extrinsic curvature. For $G=O(r^{-6})$, we find that with
$\psi=O(r^{-1})$ the non-principal terms in (\ref{ellBYpuncture}) are
$G\psi^{-7}=O(r)\in C^0$, i.e.\ continuous, and a solution $u$ of
(\ref{ellBYpuncture}) is therefore expected to be twice differentiable
at the puncture, $u\in C^2$ \cite{BraBru97}.  This is sufficient for
second order finite differences, but we have to expect numerical
issues for higher order approximations. In practice some higher order
difference schemes can be applied to improve accuracy,
e.g.\ \cite{GalBruCao10}.  Furthermore, a coordinate transformation can
raise the differentiability at the puncture to $C^\infty$, so a
pseudospectral method can show exponential convergence
\cite{AnsBruTic04}.  Depending on the extrinsic curvature, there may
be issues with the uniqueness (and/or existence) of solutions to the
Hamiltonian constraint. Essentially, for $K\neq0$ there is no general
theorem for existence and uniqueness of the full set of constraints in
the asymptotically flat setting, but given a concrete choice of
$K_{ij}$, some statements can be made \cite{DaiFri01}.

\subsection{Trumpet puncture} \label{Trumpet puncture}

A key difference when working with non-Bowen-York type extrinsic
curvature and/or changing the boundary conditions of the Hamiltonian
constraint is the possibility that the singularity of the conformal
factor at the puncture changes. For wormholes $\psi \sim 1/r$, while
for standard trumpets $\psi \sim 1/\sqrt{r}$. Geometrically, a trumpet
is one half of a wormhole. The puncture point $r=0$ corresponds to 
a finite value of the Schwarzschild radial coordinate $R(r)$, $R(0)=R_0$.

Consider the 1+log trumpet for the Schwarzschild spacetime 
that arises in puncture evolutions with the
moving puncture gauge, where $R_0\approx1.312M$. The extrinsic
curvature terms in the Hamiltonian constraint (\ref{ham1}) assume a
finite value, i.e.\ $G=0$ and $F = (K_{ij}K^{ij} - K^2)/8$,
$F_0>0$, and in particular $K_0\approx0.3009 M^{-1}$. If we make the 
ansatz that $\psi$ behaves like some power of $r$ at the puncture,
and that $F$ approaches a constant $F_0$,
then by simple power-counting using (\ref{ham1}),
\beq
  \psi\sim r^q, \quad \Delta \psi \sim \psi^5 
  \Rightarrow r^{q-2} \sim r^{5q}
  \Rightarrow q = -\frac{1}{2}. 
\eeq
Detailed calculations confirm this behavior \cite{HanHusPol06}.

Although not trivially given as in the case of a Schwarzschild
wormhole, we can compute
\beq
  \psi = \psi_{trumpet}, \quad F = F_{trumpet}
\eeq
semi-analytically at the cost of a one-dimensional integration, see \cite{Ohm08,Bru09} 
and section \ref{Single 1+log trumpet}.
For the numerical computation of trumpet data, we therefore
do have a similar starting point as in the case of wormhole data, i.e.\
we are given the Schwarzschild case. The question is how we can extend
the Schwarzschild trumpet to the spinning/moving case and the case of
multiple punctures. The catch is that now the Schwarzschild solution
does not drop out trivially when making the ansatz
\beq
  \psi = \psi_{trumpet} + u,
\eeq
since $\Delta\psi_{trumpet}=-F_{trumpet}\psi^5_{trumpet}$ does not vanish but rather gives a
curvature term. 
Consider for example two Schwarzschild punctures (no spin, no
momentum) at different locations with two solutions
\beq
	\Delta \psi_\n + F_\n \psi^5_\n = 0.
\eeq
If we set 
\beq
\psi = \psi_\na + \psi_\nb + u,
\label{puncturetrick2}
\eeq
then the Hamiltonian constraint becomes
\bea
  	\Delta u &=& \Delta \psi - \Delta\psi_\na - \Delta\psi_\nb
\nonumber\\
	&=& - F (u+\psi_\na+\psi_\nb)^5 
            + F_\na \psi^5_\na + F_\nb \psi^5_\nb.
\nonumber\\
&&\label{Du}
\eea
For wormhole punctures with Bowen-York extrinsic curvature, the
corresponding right-hand-side of (\ref{Du}) would be non-singular and the
coefficient would vanish
sufficiently fast at the punctures such that $u\in C^2$. However, 
for trumpet punctures the leading order behavior is determined
by the $\psi^5\sim r^{-5/2}$ terms.
We assume that not only the $F_\n$ are non-zero, but
also that the combined $F$ is non-zero at the puncture, and that there
are no unexpected cancellations in (\ref{Du}). Then we conclude that
$u$ is as singular as $\psi$, i.e.\ $u \sim r^{-1/2}$, and we have not
gained regularity for the numerical solution. 
This means that the additive correction of the original puncture trick
as given in (\ref{puncturetrick2}) is not sufficient for 1+log
trumpets. Rather, we should look for a different way to handle the
$r^{-1/2}$ singularity at the puncture.

It is possible to move parts of the singularity into the analytic part
of the conformal factor, as done for maximal trumpets in
\cite{HanHusOMu06,HanHusOMu09}, and this approach could be attempted
for 1+log trumpets as well. As an alternative we considered a
multiplicative puncture trick, $\psi = \psi_S \chi$, where $\psi_S$
contains the singular part. Numerical experiments with this ansatz
were successful, but the accuracy was several orders of magnitude
lower than for (\ref{psiq}). We did not pursue this option
further. Let us mention that \cite{AnsBai13} had some success with
transforming the radial coordinate by $s=\sqrt{r}$.

The proposal in the present work is to insert (\ref{psiq}), $\psi=f^p$, into the
Hamiltonian constraint (\ref{ham1}), which with (\ref{Deltafp}) leads to
\beq
 \Delta f -(p-1) \frac{(\nabla f)^2}{f}+\frac{F}{p} f^{4p+1}+\frac{G}{p} f^{-8p+1}=0.
\label{Hamiltonregular}
\eeq
We have normalized the principal part, as is customary and often
advantageous for a numerical implementation.
As examples we consider a wormhole puncture with BY extrinsic curvature 
for $p=-\frac{1}{2}$, $f\simeq r^2$,
\beq \textstyle
   \Delta f - \frac{3}{2} \frac{(\nabla f)^2}{f} - 2 G f^5 = 0,  \label{psim2}
\eeq
and the 1+log Schwarzschild trumpet for $p=-\frac{1}{2}$, $-\frac{1}{4}$, $-\frac{1}{8}$
and $f\simeq r$, $r^2$, $r^4$,
\bea
\textstyle \Delta f - \frac{3}{2} \frac{(\nabla f)^2}{f} - 2 F f^{-1} - 2 G  f^{5} &=& 0,
\label{psim2tr}
\\
\textstyle \Delta f - \frac{5}{4} \frac{(\nabla f)^2}{f} - 4 F - 4 G f^{3}  &=& 0, 
\label{psim4} 
\\
\textstyle \Delta f - \frac{9}{8} \frac{(\nabla f)^2}{f} - 8 F f^{1/2}  -8 G f^{2} &=& 0, 
\label{psim8}
\eea
respectively. The leading order behavior of the terms with derivatives is
therefore $r^{-1}$, $1$, or $r^2$.
In all cases the 
division of a numerical derivative by $f$ in $\frac{(\nabla f)^2}{f}$ 
may or may not be numerically
tricky. Furthermore, we have to discuss the regularity of the terms
$F f^{4p+1}$ and $G f^{-8p+1}$. A difference between maximal and 1+log trumpets
is $F\neq0$ ($K\neq0$). While $G f^{-8p+1}$ vanishes sufficiently rapidly for our 
choices of $p$, $F f^{4p+1}$ contributes at the same leading order
as the derivative terms. The different values of $p$ are chosen to examine 
whether increasing the smoothness of $f\simeq r^{-1/(2p)}$ near the puncture helps,
but it turns out that increasing the order in $r$ can be detrimental.
Eqn.\ (\ref{psim4}), $f\simeq r^2$, was one of the examples considered 
in \cite{Gun10a}, since with $-4Ff^0=K^2/3\simeq const.$ the non-vanishing $K$ enters
in a rather simple manner.
The first impression that $F/f$ in (\ref{psim2tr}) will 
cause problems when computing the right hand side turns out to be wrong.
In our examples, $p=-\frac{1}{2}$ leads to the most accurate results. 

\section{Solving the Hamiltonian constraint with regular conformal factor}
\label{RewHam}

\subsection{Numerical method}

To test the method we use a new implementation of the single domain
pseudospectral code of \cite{AnsBruTic04}, which is described in
\cite{Die12} in more detail. The goal was not to obtain exponential
convergence, but rather to employ an existing, efficient method that
can be expected to show polynomial convergence even in the presence of
the trumpet singularity. Since the required grid size turns out to
be rather small (a 3d or 2d grid with no more than ten thousand points total), we can use a direct
linear matrix solver inside a Newton-Raphson iteration.

For the single puncture we introduce compactified spherical coordinates 
$(A,\theta,\varphi)$, where we have compactified according to 
\beq
A=\left( 1+\frac{m}{2 r}\right)^{-1}. \label{singdom}
\eeq
The computational domain consists of a Chebychev grid in the radial direction 
and two Fourier grids for the angular quantities, which we denote as a CFF grid
(an alternative to spherical harmonics, e.g.\ \cite{Mer73,For98,Bru11}).
For the collocation points we choose the staggered Chebychev grid that does not include
points at the boundary, i.e.\ the 
zeros of the Chebychev polynomials $T_{n_A}(1-2A)$,
$\sin(n_\theta \theta)$, and $\sin(n_\varphi \varphi)$, where 
$n_A,n_\theta,n_\varphi$ denote the number of grid points in each direction. 
The puncture is located on the (two-dimensional) $A=0$ boundary. 
This improves the convergence behavior of the spectral method, because no kink or pole is located 
in the interior of the grid. As a consequence of the staggering, we avoid 
outright division by zero at the puncture, although some numerical issues remain
as the points cluster quadratically near $A=0$.  On the other hand, we can not impose a Dirichlet 
boundary condition at infinity trivially since the grid is staggered there as well. 
We implement the outer boundary at $A=1$ by extrapolation (which in our case gives more
accurate results then the variable substitution described in \cite{AnsBruTic04}).

For two punctures we introduce compactified prolate spheroidal coordinates, 
in our notation $(A,B,\varphi)$, with the inverse coordinate transformation 
to Cartesian coordinates given by
\bea
x &=& b \frac{A^2+1}{A^2-1} \frac{2 B }{1+B^2} ,\\
y &=& b \frac{2A} {1-A^2} \frac{1-B^2}{1+B^2} \cos (\varphi ) ,\\
z &=& b \frac{2A} {1-A^2} \frac{1-B^2}{1+B^2} \sin (\varphi ) .
\eea
This coordinate transformation was introduced for wormhole initial
data in~\cite{AnsBruTic04}. It was shown that there is a 
correction $u$ that is $C^\infty$ at the punctures.  The computational
domain is built up of either a Chebychev-Chebychev-Fourier (CCF)
or a Chebychev-Fourier-Fourier grid (CFF).
The grid points for the CCF grid are the
zeros of $T_{n_A}(1-2A)$, $T_{n_B}(-B)$, and $\sin(n_\varphi
\varphi)$. The radial-type coordinate is again denoted by $A$ with
$A=1$ at spatial inifity. The coordinate $B$ runs from $-1$ to $1$
and the two punctures are located at $A=0,B=\pm1$. Thus, the black
holes are at (one-dimensional) edges of the grid. As for the single
puncture we extrapolate to spatial infinity.
For the CFF grid we introduced a double covering in $B$, 
where $B$ runs from $-2$ to $2$, and 
the first black hole is at $A=0,B=0$ and the second at $A=0,B=\pm2$. The 
grid points are the zeros of $\sin(\frac{\pi}{2} n_B B)$.

\subsection{Single wormhole}

For a single wormhole with a regular conformal factor, we have $K=0$ and thus 
$F=0$, while $\bar{A}_{ij} \bar{A}^{ij}$ is defined by the Bowen-York curvature 
\begin{equation}
 \begin{aligned}
\bar{A}_{ij}= \frac{3}{2r^2} ( n_i P_j+n_j P_i+ n_k P^k (n_i n_j -\delta_{ij})  \\
- \frac{3}{r^3} (\epsilon_{ilk} n_j + \epsilon_{jlk} n_i ) n^l S^k ), 
\end{aligned}
\end{equation}
where $P^i$ is the momentum and $S^k$ the spin of the black hole.
The outward-pointing unit radial vector is denoted by $n^i$.
As described in the introduction, wormhole initial data are widely
used in numerical relativity, but solving the Hamiltonian constraint
for wormholes considering a regular factor is a novel idea, see also
\cite{Bau12a}. 

\begin{figure}[tpb]
\includegraphics[width=0.5\textwidth]{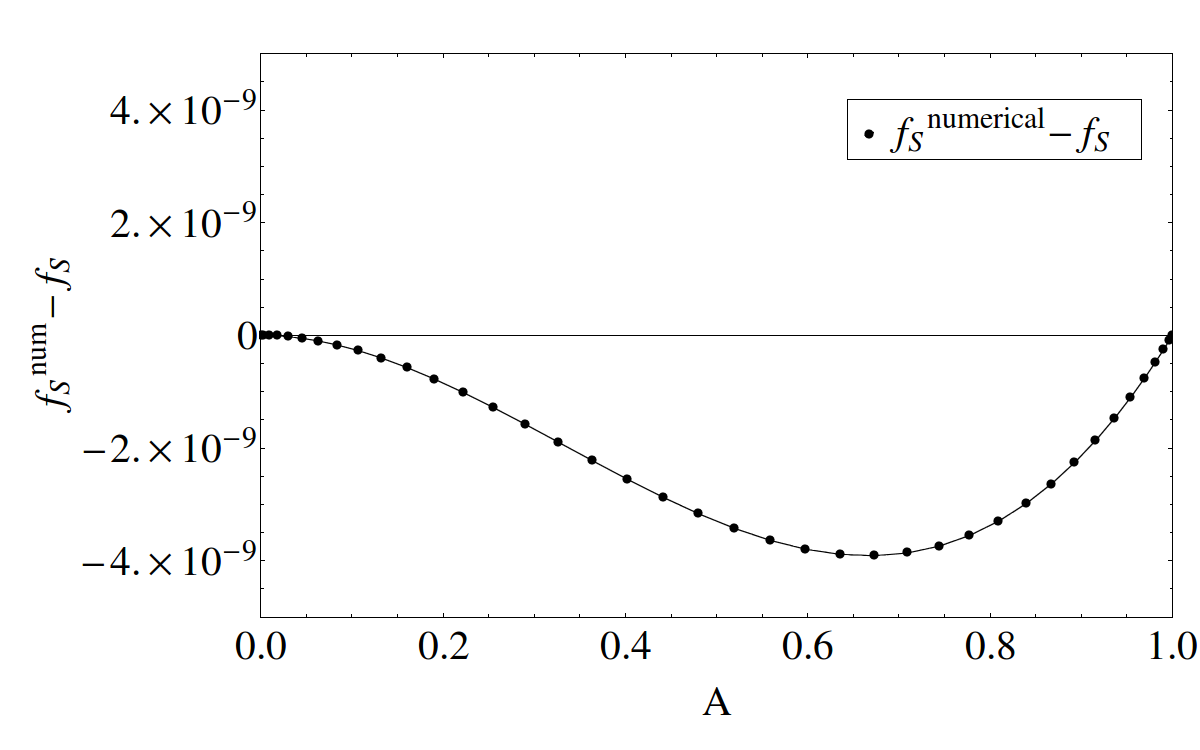} 
\caption{
 Single wormhole, Schwarzschild solution. 
 Difference between the numerical result for the regular 
 conformal factor $f$ obtained with the pseudospectral code 
 for $n_A=40$, $n_\theta=2$, $n_\varphi=1$
 and the analytical solution. 
}
\label{Figsingwh}
\end{figure}

For the computation of the regular factor $f$ we have to define two
boundaries. One is the outer boundary, where we set $f|_\infty=1$, the
other refers to the puncture point. In the standard puncture method
the inner boundary condition is not needed, because the mass of the black hole is
imposed via $\psi_S=1+\frac{m}{2r}$. For the regular conformal factor, 
the mass of the black hole is not fixed, so we find an entire branch of solutions,
making the numerical code fail (the linear problem is
underdetermined) unless we impose the mass of the wormhole by
hand. This is to be expected, see Sec.~\ref{WormPunc}.
For $p=-1/2$, we impose a condition on the second derivative of $f$ at
the puncture.  We set $\partial_{A}^2f=2$, so that $f \sim A^2$ near
the puncture. It is straightforward to see that this also holds for BY
extrinsic curvature. The mass scale is then given by $m$ in the definition of
$A$, see (\ref{singdom}).
With the correct boundary condition, the method solves the Hamiltonian
constraint for the Schwarzschild wormhole for vanishing extrinsic
curvature without problems and obtains a numerical approximation to
$f_S=\psi_S^{-2}$ with rather smooth numerical error, see
Fig.~\ref{Figsingwh}. In particular, there are no numerical artefacts
at the puncture.  We have confirmed that the method also works for
single wormholes with BY extrinsic curvature for spin and linear
momentum.

\subsection{Single 1+log trumpet} \label{Single 1+log trumpet}

When solving (\ref{Hamiltonregular}) for a single trumpet, we can
compare our results with the semi-analytical known solution. 
We integrate the 1+log condition as in \cite{Bru09}, obtaining
with $S=1/R$
\bea
F(S,\alpha) &\equiv& \alpha^2 -1+2S-C e^\alpha S^4=0, \\
\frac{\partial F}{\partial S} &=& 2 - 4 C e^\alpha S^3, \\
\frac{\partial F}{\partial \alpha} &=& 2 \alpha -C e^\alpha S^4.
\eea
We compute $S(\alpha)$ with the Newton-Raphson method, and from 
\beq
\frac{dS}{dr}=  -\frac{S(r)}{\alpha(S(r))}{r}
\eeq
we obtain
\beq
\psi^{-2}(r)=\frac{r}{R(r)}. \label{semianamp1}
\eeq

\begin{figure}[tpb]
\includegraphics[width=0.5\textwidth]{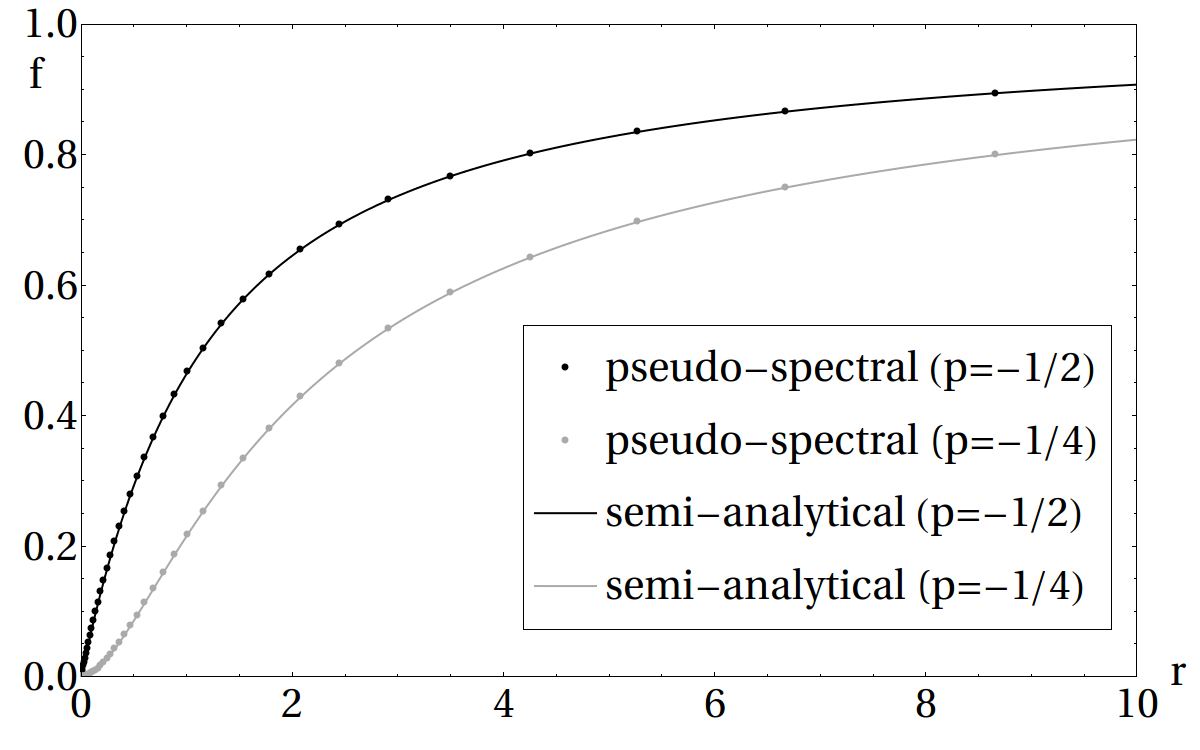} 
\includegraphics[width=0.5\textwidth]{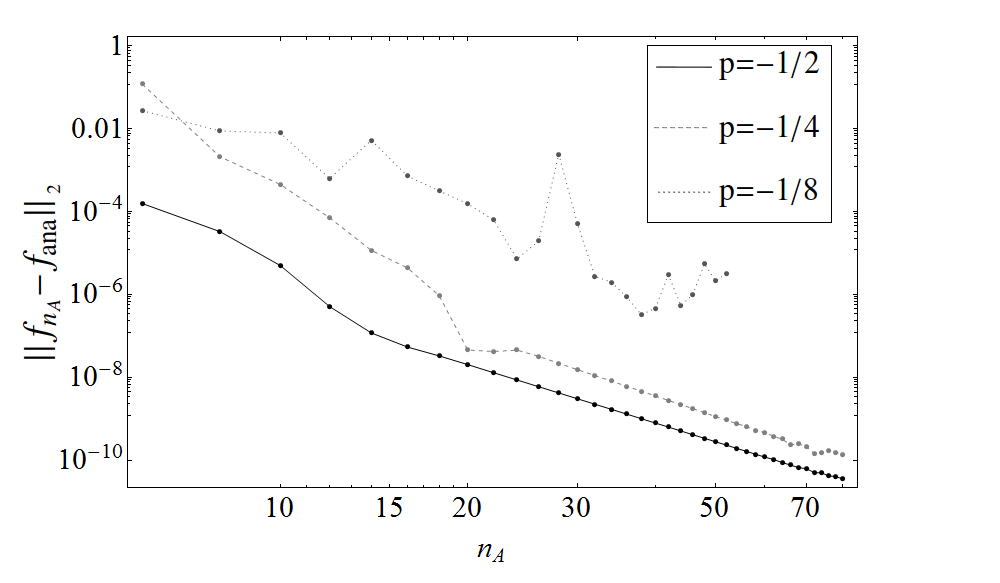}
\caption{
Single 1+log trumpet, Schwarzschild solution.
Comparison between the numerical and the semi-analytical solution 
(upper panel) for different choices of $p$, and a log-log plot of the convergence in the 
$l_2$-norm of the difference between the numerical and the semi-analytical solution 
(lower panel). 
In the upper panel $n_A=50$, and in both panels $n_\theta=2$, $n_\varphi=1$.
For large $n_A$ there appears to be polynomial convergence of order $5\pm0.5$.
}
\label{mp1}
\end{figure}

Fig.~\ref{mp1} is a comparison between the results achieved by our
single domain pseudospectral code and the semi-analytical solution
(\ref{semianamp1}). The upper panel shows that our code finds the
correct solution. This was not clear from the beginning because of
uniqueness issues revealed by \cite{Gun10a}. But since we are
interested in $\psi,f>0$ a power series expansion can be used to
visualize why it is likely to find the correct solution with an
appropriate initial value close to the analytical solution
\cite{Die12}.

According to the lower panel of Fig.~\ref{mp1} the method does not
achieve exponential convergence, except perhaps for small $n_A$, but
this behavior was expected.  On the one hand, we have regularity
issues at the puncture, and on the other hand, 
there may be
logarithmic terms at spatial infinity. Both effects lead to
polynomial convergence of the pseudospectral code, which should explain
the observed polynomial order of about 5.
More important is the qualitative difference between the choices
of $p$. As mentioned in Sec.~\ref{Trumpet puncture}, regarding the regularity 
of $F f^{4p+1}$ one might expect that $p=-1/2$ can cause difficulties
because of the division by $f$. However, Fig.~\ref{mp1} indicates that 
$p=-1/2$ gives more accurate solutions than $p=-1/4$, while $p=-1/8$ is significantly
less accurate than both these choices.
The reason for this could be that a linear function does not require as 
much numerical resolution as  
a quadratic or quartic function, which explains why 
the results for $p=-1/2$ are better than for the others. 
Although $\frac{(\nabla f)^2}{f}$ is analytically zero at the puncture
for $p=-1/4$ or $p=-1/8$, there are indications that with increasing
exponent of $r$ some accuracy is lost.

\begin{figure}[tpb]
\includegraphics[width=0.5\textwidth]{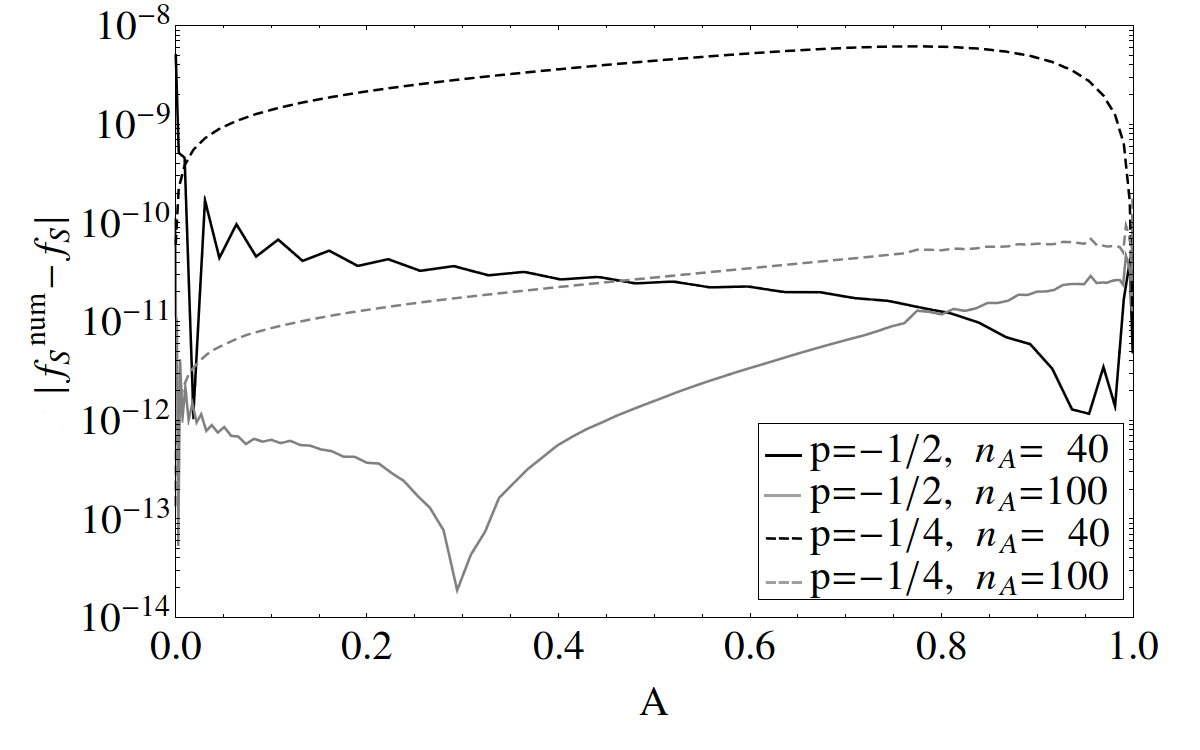} 
\caption{\
Single 1+log trumpet, Schwarzschild solution.
Absolute difference between the pseudospectral solution and the 
semi-analytical result versus $A$ for different values of $p$ and $n_A$.
}
\label{single1log}
\end{figure}

Fig.~\ref{single1log} shows the logarithm of the error with respect to
the semi-analytical solution for $p=-1/2$ and $p=-1/4$ versus $A$.
Note the numerical noise for $p=-1/2$ near the puncture ($A=0$), which
is likely due to the $1/r$ behavior of the Hamiltonian constraint,
with additional issues near infinity ($A=1$).  For very high
resolution, $n_A=100$, there are numerical issues both at the puncture
and at infinity even for $p=-1/4$. However, this resolution is
significantly higher than in Fig.~\ref{mp1}, and various numerical
round-off effects can make the error larger than for lower
resolutions. Overall, an error on the order of $10^{-10}$ is certainly
acceptable for our purpose.

\subsection{Multiple trumpets}

Solving the Hamiltonian constraint for a regular conformal factor enables us to
solve for the first time the Hamiltonian constraint for binary
1+log {\em trumpet} data.  
In contrast to the original puncture trick, where the generalization from one
1+log trumpet to two trumpets fails, using inverse powers of the conformal factor in the
Hamiltonian constraint works without additional difficulties.
Recall that our strategy is to postpone the solution of the momentum constraint, but
we can still evaluate different approximation strategies for specifying the extrinsic
curvature.

We decompose the extrinsic curvature according to $F= -K^2/12$ and
$G=\bar{A}_{ij} \bar{A}^{ij}/8$ as we did it for the single trumpet.
This approach improves the stability of the code and decreases the
residuum. Similar to the single trumpet we achieve only polynomial
convergence.
As a first ansatz we consider the following possibilities to specify $F$ and $G$ 
as a superposition of two single 1+log trumpets:
\bea
F&=&-\frac{1}{12} (K_{(1)}^2+K_{(2)}^2), \label{superpos1}\\
F&=&-\frac{1}{12} (K_{(1)}+K_{(2)})^2, \label{superpos2}\\
G&=& \frac{1}{8}  ({\bar{A}^{(1)}}_{ij} {\bar{A}_{(1)}}^{ij}
+{\bar{A}^{(2)}}_{ij} {\bar{A}_{(2)}}^{ij}), \label{superpos3}\\
G&=& \frac{1}{8}  ({\bar{A}^{(1)}}_{ij} +{\bar{A}^{(2)}}_{ij})
 ({\bar{A}_{(1)}}^{ij}+{\bar{A}_{(2)}}^{ij}),  \label{superpos4}\\
G&=& \frac{1}{8}  (A^{(1)}_{ij} A_{(1)}^{ij}
+A^{(2)}_{ij} A_{(2)}^{ij}) \psi_0^{-12},\label{superpos5}
\eea
where $\psi_0= \psi^{(1)}_0+\psi^{(2)}_0-1$ and $\psi^{(i)}_0$ are 
the solutions for single trumpets. We can use these simple superpositions since 
the source terms behave for large $r$ like $F \sim r^{-8}$ and $G \sim r^{-6}$. 
Thus, for sufficient separation the effect of the superposition is negligible.
Although we could add additional momentum, say of the BY type, as specified
the two punctures are in an axisymmetric, head-on configuration, 
in which they are approximately at rest.

\begin{figure}[tpb]
\includegraphics[width=0.5\textwidth]{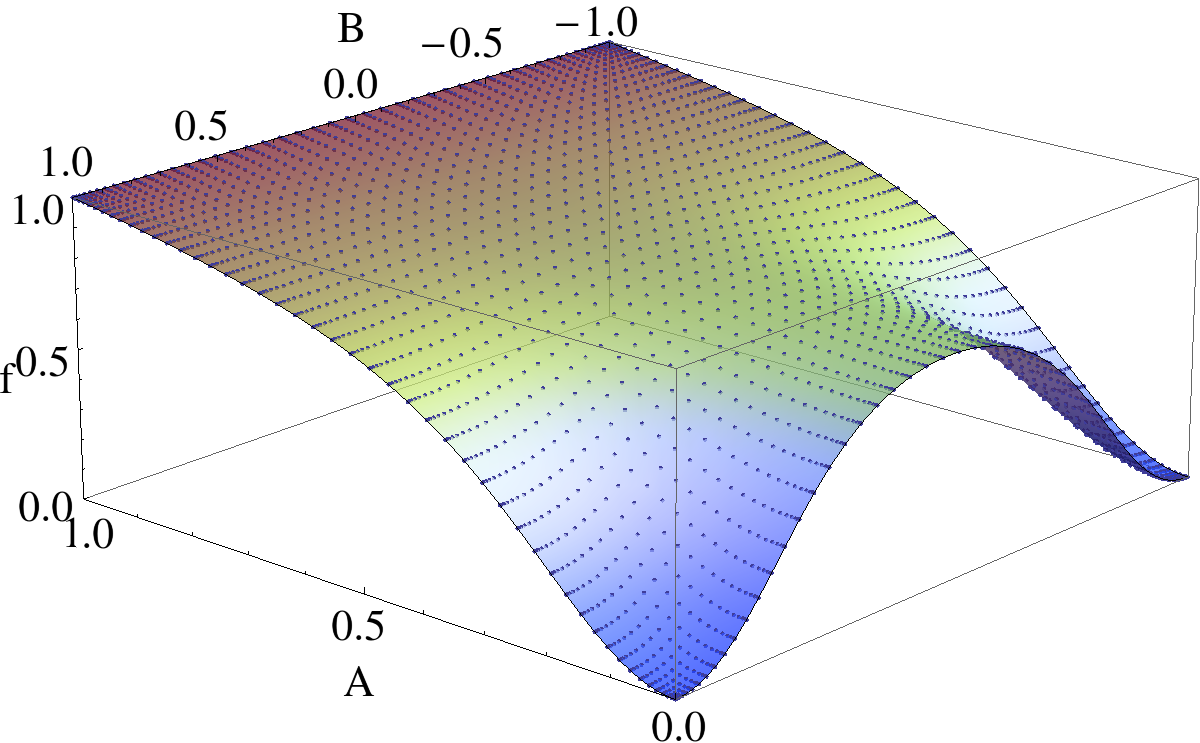} 
\includegraphics[width=0.5\textwidth]{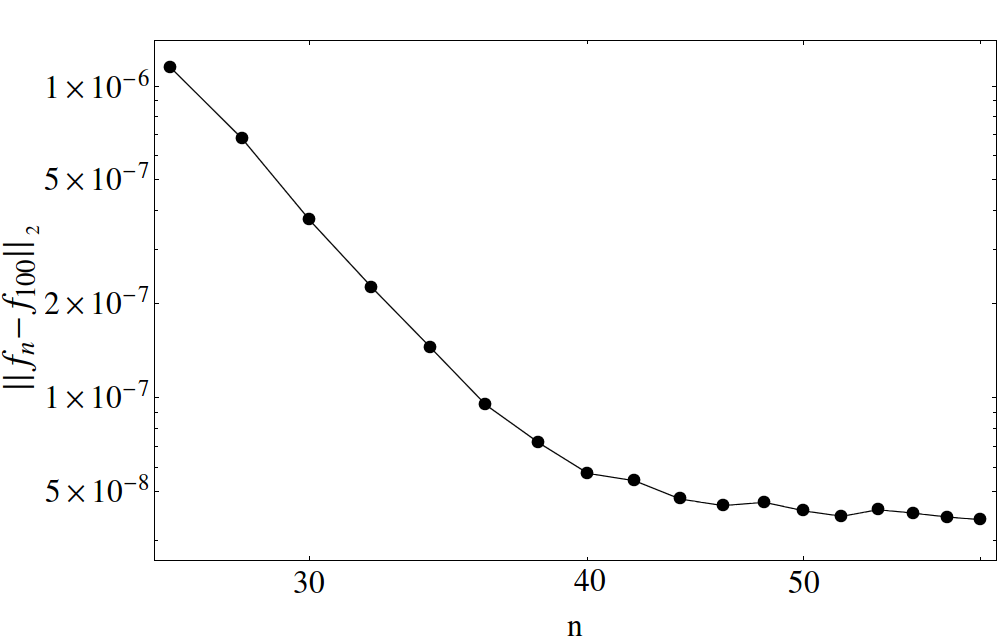} 
\caption{
Two 1+log trumpets, axisymmetric head-on case. The conformal factor $f$ is shown for
separation $d=12m$ and masses $m_1=m_2=m$ using $p=-1/2$,
(\ref{superpos1}), and (\ref{superpos5}) (upper panel).
Also shown is the convergence of the error with $n=n_A=n_B$,
for which we compute the $l_2$-norm of the difference to the solution for 
$n=100$ (lower panel). 
}
\label{mp2}
\end{figure}

We tested all possible combinations of
(\ref{superpos1})-(\ref{superpos5}) and the code produces reasonable,
and despite minor differences, similar results.  As a particular
example we consider two black holes with mass $m_1=m_2=m$ at a
separation of $d=2b=12m$. In Fig.~\ref{mp2} we show for $p=-1/2$ the
regular factor $f$ using (\ref{superpos1}) and (\ref{superpos5}). We
set $n=n_A=n_B$, and $n_\phi=1$ because of axisymmetry.

\begin{figure}
\includegraphics[width=0.5\textwidth]{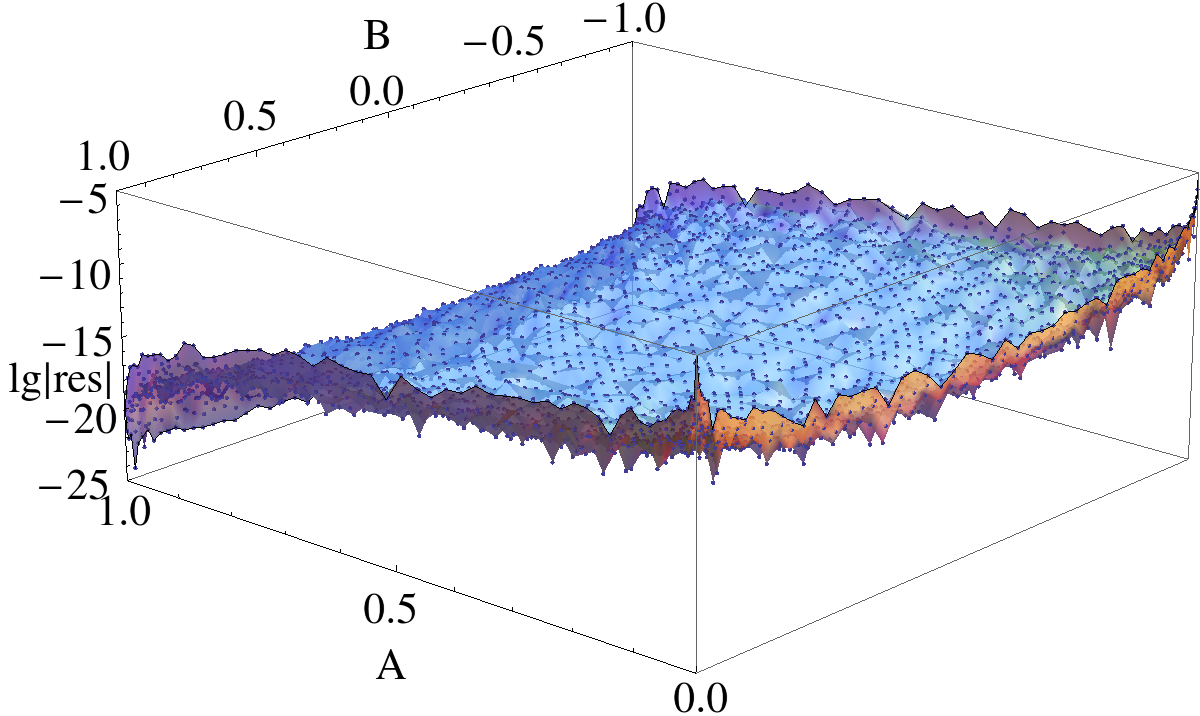} 
\includegraphics[width=0.5\textwidth]{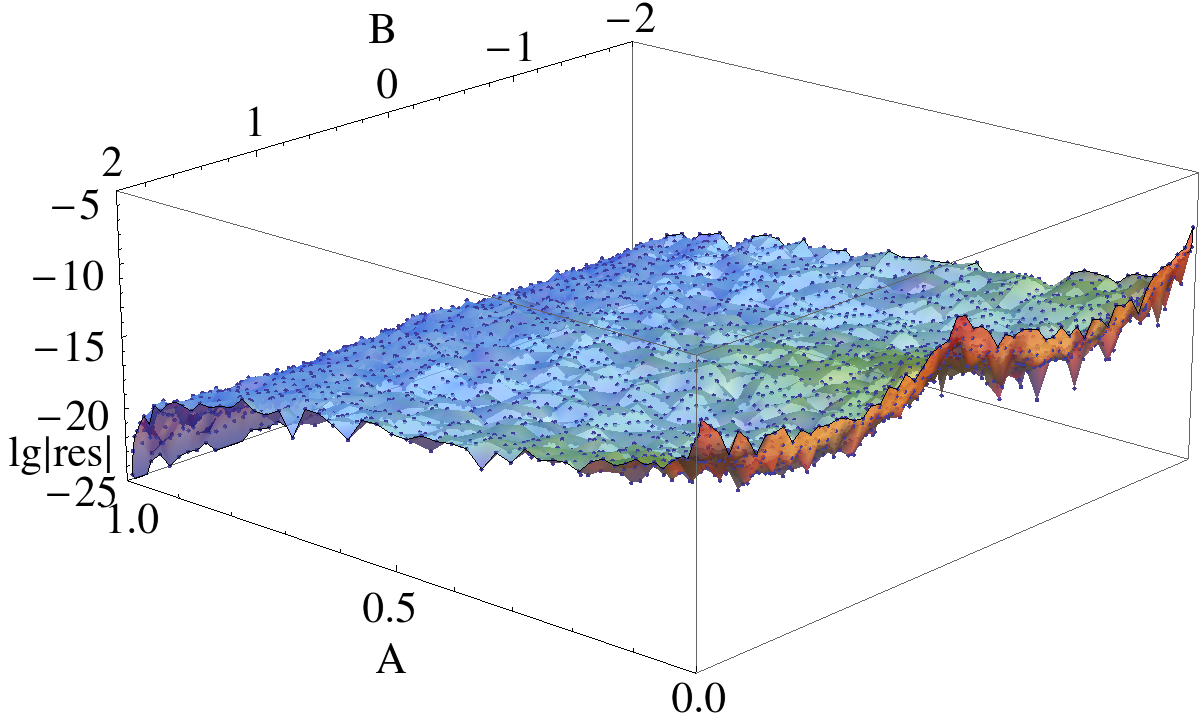} 
\caption{Two 1+log trumpets as in Fig.~\ref{mp2}. 
Logarithm of the absolute value of the residuum for two 1+log trumpets 
with $n_A=n_B=60$, $n_\varphi=1$. The upper panel shows the result for the CCF configuration, 
the lower panel for CFF. Along the axis, which includes the punctures,
the residuum is several orders of magnitude smaller for CFF than for CCF.}
\label{mp2res}
\end{figure}

In the cases we tried, the CCF method (similar to \cite{AnsBruTic04})
was more robust than the CFF method, i.e.\ we could solve for a larger
range of parameters with less grid points. For this reason we prefer
the CCF grid and unless stated otherwise present results with this
setup.  On the other hand, the residuum for the CFF grid was smaller
than for CCF, see Fig.~\ref{mp2res}.

One of the features of 1+log trumpet initial data 
is the reduction of gauge dynamics. This reduction can be improved 
by a special choice of the initial shift and initial lapse for our evolutions, 
which is similar to choosing a pre-collapsed lapse for wormhole punctures.
On the one hand, we can use a simple superposition 
\bea
\alpha&=& \alpha_{(1)} + \alpha_{(2)} -1  + \epsilon \label{alphasup},\\
\beta^i&=& \beta^i_{(1)} + \beta_{(2)}^i, \label{betasup}
\eea
where $\epsilon$ is a small parameter to ensure that $\alpha>0$ at the punctures. 
A coordinate dependent choice for $\epsilon$ is in principle possible, but
was not tried in our investigations. 

On the other hand, we can obtain lapse and shift by solving the
corresponding equations of the 
conformal thin-sandwich (CTS)
decomposition \cite{Yor99},

\bea
(\Delta_L \beta )^i &=& (L \beta)^{ij} \partial_j 
\ln (\alpha \psi^{-6}) +\frac{4}{3} \alpha \partial^i K, \label{CTSbeta}\\
 \Delta (\alpha \psi) &=& \alpha \psi \left( \frac{7}{8} \psi^{-8} 
\bar{A}_{ij} \bar{A}^{ij}  + 
\frac{5}{12} \psi^4 K^2  \right)  \nonumber  \\
 &+& \psi^5 \beta^i \partial_i K, \label{alphapsi}
\eea
where we use (\ref{alphasup}) and (\ref{betasup}) for the computation of the source terms in the right hand side  of the equations.
Additionally, we have set $\partial_t K= \partial_t \bar{g}_{ij}=0$ and denoted the vector Laplacian by $\Delta_L$, 
while $( L \beta )^{ij}= \partial^i \beta^j +\partial^j \beta^i - \frac{2}{3} \bar{g}^{ij} \partial_k \beta^k$. 
Both ansatzes lead to approximately the same behavior in black hole
evolutions. In Sec.~\ref{Evo} we discuss
results for (\ref{alphasup}) and (\ref{betasup}), because they are
easier to obtain and the difference between those results and results obtained 
with (\ref{CTSbeta}) and (\ref{alphapsi}) seems to be minor.

\section{Evolution}
\label{Evo}

The initial data are evolved with the BAM code
\cite{HusGonHan07,BruGonHan06,BruTicJan03} using the BSSN evolution
scheme with $\tilde{\Gamma}$-driver shift and
1+log-slicing.

\subsection{Constraint violations}

Evolutions of our trumpet data allows us to examine the effect of
working with an ad hoc ansatz for the extrinsic curvature rather than
solving the momentum constraint. Since we are not using maximal
slicing, a simple superposition of the extrinsic curvature terms for
single black holes is not a solution for the momentum constraint.  The
investigation of (\ref{superpos1})-(\ref{superpos5}) revealed no
significant difference in the constraint violation between
(\ref{superpos1}) and (\ref{superpos2}).  But minor differences
depending on the superposition of $G$ occurred, where
(\ref{superpos5}) leads to the smallest constraint violation in our
simulations.  For evolutions, we define the initial extrinsic
curvature by
\beq
\bar{A}_{ij}  = \bar{A}^{(1)}_{ij}+\bar{A}^{(2)}_{ij},  
\label{extCurv}
\eeq
and raise indices with $\bar{g}^{ij}$ as usual,  
while in (\ref{superpos3})-(\ref{superpos5}) indices are raised
with the single puncture metric.
A priori we do not know the momentum constraint violation and from an
analytical point of view it is quite debatable how well our ansatz
will work. However, we find that the momentum constraint violation is
dominated by the evolution itself and not by the inaccuracies of the
initial data.

Eqn.\ (\ref{extCurv}) may lead to a large violation 
of the momentum constraint near the puncture. For simplicity we consider 
the initial guess for $\psi_0 = \psi_0^{(1)} + \psi_0^{(2)} -1$.  We assume
$\psi_0^{(2)}-1=\xi \ll \psi_0^{(1)}$ near the first puncture. 
Then, the momentum constraint using the CTT decomposition and 
the conformal factor $\psi_0$ turns out to be
\beq
\begin{aligned}
\partial_j   \bar{A}^{ij} - \frac{2}{3} ( \psi_0^{(1)}+\xi)^6 
\bar{g}^{ij} \partial_j K \approx  \\
\qquad - 4 \xi (\psi_0^{(1)})^5  \bar{g}^{ij} \partial_j
K_{(1)} \neq 0.\label{momconst}
\end{aligned}
\eeq
At the first puncture $\psi_0^{(1)} \rightarrow \infty$, which leads to a 
divergent constraint violation at the position of the first puncture. 

\begin{figure}[tpb]
\centering
\includegraphics[width=0.45\textwidth]{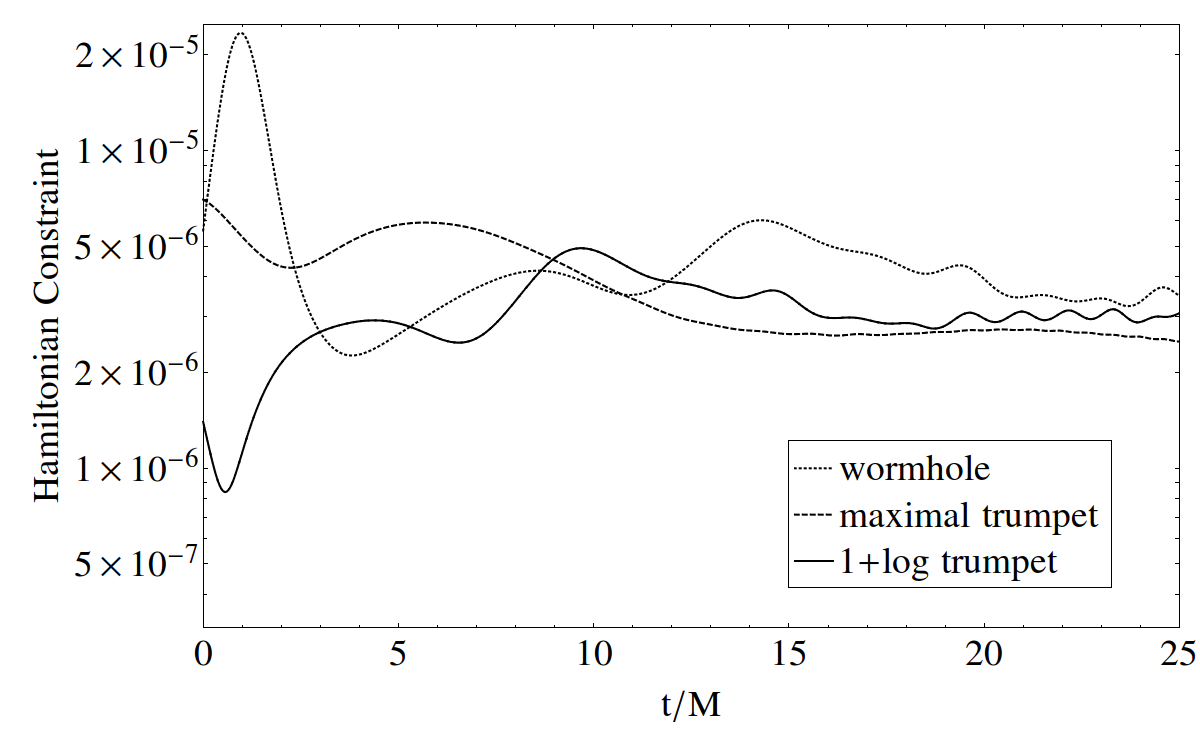}
\includegraphics[width=0.45\textwidth]{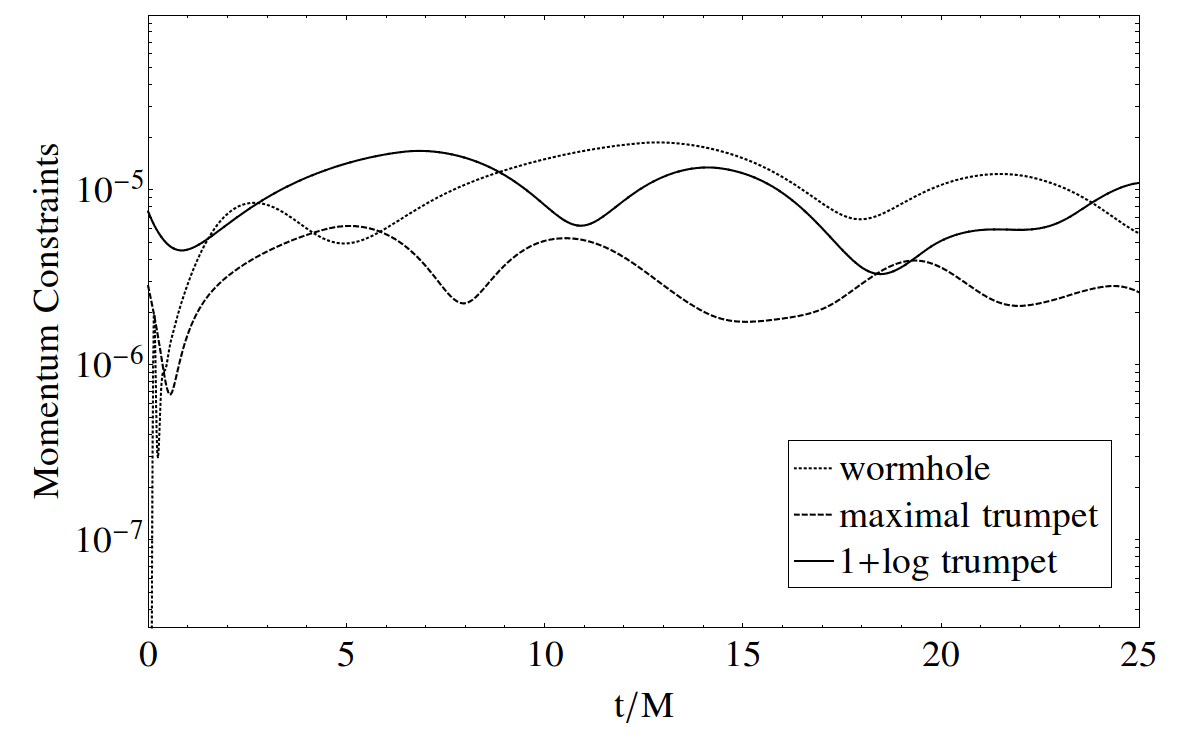}
\caption{
  Evolution of two black holes at the beginning of a head-on collision.
  Shown is the Hamiltonian constraint violation (upper panel) and norm of the
  momentum constraints (lower panel) for three types of initial data. 
}
\label{Figconstraints}
\end{figure}

However, evolutions with the puncture method are able to handle
certain intrinsic regularity issues of the punctures, and in practice
this is also the case for the momentum constraint violating initial
data that we constructed. The question is how large the constraint
violations are for evolutions of the approximate 1+log trumpet data
compared to standard wormhole evolutions.  As indicated by
Fig.~\ref{Figconstraints}, the constraint violation has comparable size,
which suggests that the constraint violation
produced by the evolution is the leading order effect.
The black holes have a mass of $m$ each and an initial separation of
$d=20m$, while the total (ADM) mass of the system is $M$.  The $l_2$-norm
was computed on the second outermost level with a grid spacing of $1M$,
running from $-75M$ to $75M$.

\subsection{Reduction of initial gauge dynamics}

To quantify the reduction of initial gauge dynamics, we consider three
different types of initial data: maximal wormholes, maximal trumpets
and 1+log trumpets.  The results refer to an equal-mass binary black hole
simulation with an initial separation of $d=20m$, where the black
holes are located on the $x$-axis and perform a head-on collision.
Additionally, we set $\eta=0$ in the $\tilde{\Gamma}$-driver condition
for this subsection to reduce the gauge-related growth of the
coordinate distance of the apparent horizon
\cite{HanHusOhm08,BruGonHan06,MueGriBru10}. 

\begin{figure}[tpb]
\centering
\includegraphics[width=0.45\textwidth]{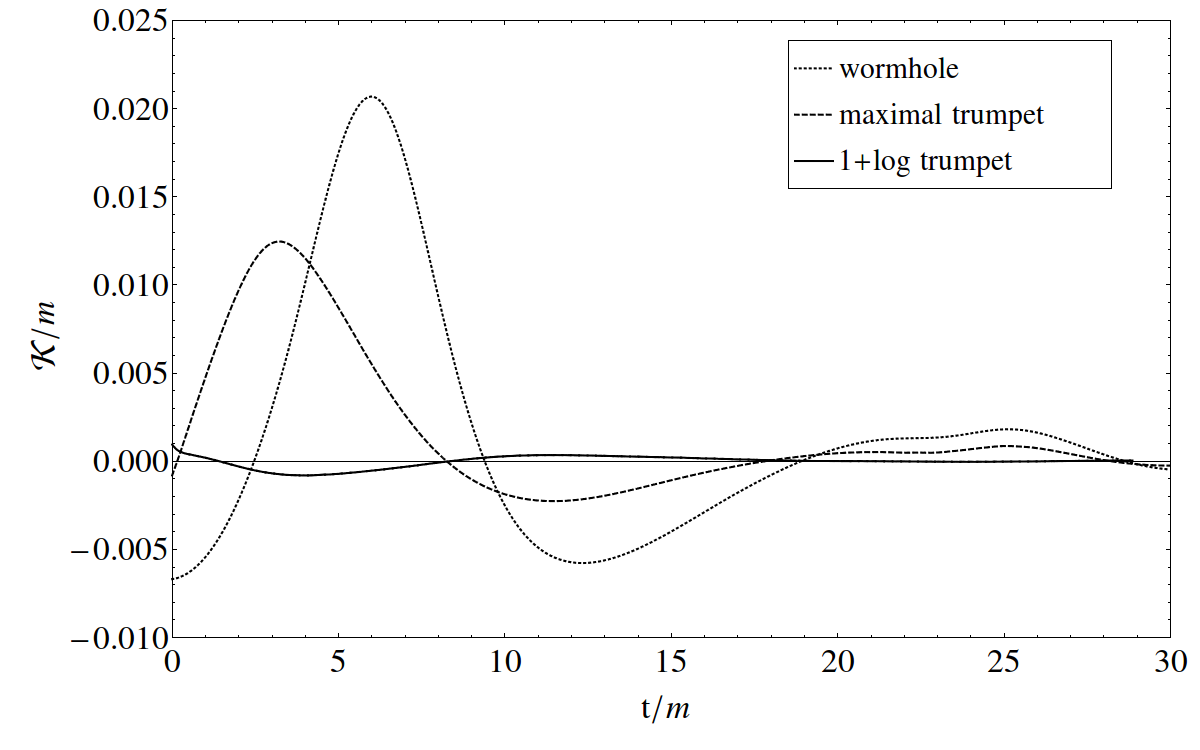}
\includegraphics[width=0.45\textwidth]{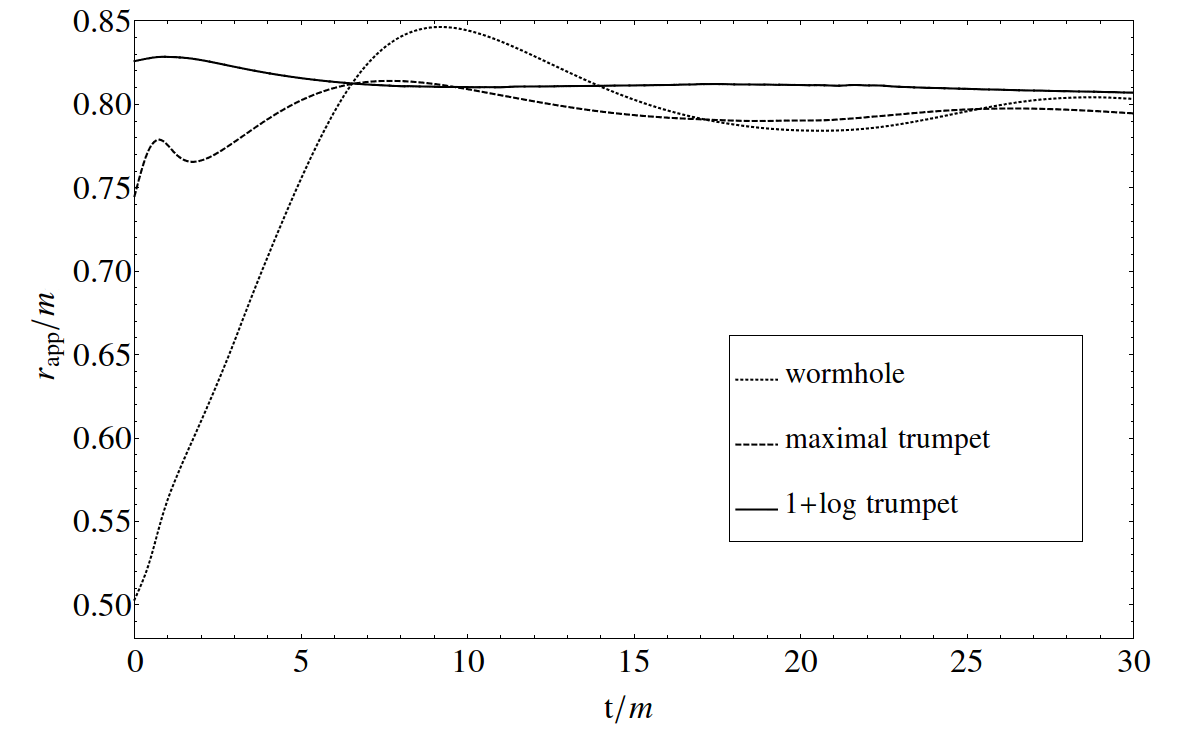}
\caption{Head-on collision of two black holes. Shown is
$\mathcal{K}$ computed with equation (\ref{deltaK}) (upper panel) 
and the mean coordinate distance of the apparent horizon (lower panel). 
We have used (\ref{alphasup}) and (\ref{betasup}) for the trumpet initial 
data and the pre-collapsed lapse $\alpha=\psi^{-2}$ for the wormholes. 
Using $\alpha=1$ and $\beta^i=0$ increases the gauge dynamics in all 
simulations.
}
\label{Figgauge}
\end{figure}

We present two quantities which show that 1+log data are the preferred
choice with respect to the initial gauge dynamics, see
Fig.~\ref{Figgauge}. Firstly, we compute the change of the trace of
the extrinsic curvature $K$ during the beginning of the
simulation. For this purpose we define the quantity
\beq
\mathcal{K}=\int\limits_{  {x_p}_{i+1}-\delta}^{{x_p}_{i+1}+\delta} K_{t_{i+1} } 
\text{d}x -\int\limits_{{x_p}_i-\delta}^{{x_p}_i+\delta} K_{t_i} \text{d}x, 
\label{deltaK}
\eeq 
where $t_{i+1}-t_i=0.125m$ and ${x_p}_i$ is the position of the
puncture for $t_i$. We choose $\delta=2 m $. Thus, we use a
one-dimensional integral to measure the growth of $K$ during the
evolution. This is reasonable because the black holes initially are nearly
spherical.

Secondly, we present the mean coordinate distance of the 
apparent horizon, 
which was already used in \cite{HanHusOMu09} to illustrate why maximal 
trumpets are a better choice for initial data than wormholes. 
In \cite{HanHusOMu09} the slicing condition 
$
\partial_t \alpha = - 2 \alpha K 
$
was used so that after about $t=10m$ the slicing was approximately 
maximal again. We will not use this equation as is, but instead include 
the shift term $\beta^i \partial_i \alpha$ on the right hand side, 
since this is normally done in black hole simulations.  

Regarding the upper panel of Fig.~\ref{Figgauge}, we conclude that the
initial change of $K$ is reduced by our 1+log trumpets. There are two reasons 
for the non-vanishing $\mathcal{K}$ found for our data. On the one hand,
there still exist small gauge dynamics at the beginning of the
simulation, while on the other hand, $\mathcal{K} \neq 0$ because of the
evolution itself.  During the evolution the linear momentum of the
black hole is increasing, which leads to a small decrease in $K$.
However, it is obvious that because of the initial change from
maximal- to 1+log-slicing there has to be a change in $K$ (from zero
to non-zero) for the wormholes and the maximal trumpets. Therefore, we
present the mean coordinate distance of the apparent horizon as a
second diagnostic of the initial gauge dynamics.
The lower panel of Fig.~\ref{Figgauge} reveals the same behavior
found before, namely the 1+log trumpet is the preferred choice to 
minimize the early gauge dynamics of the evolution. The initial dynamics of the 
horizon radius is reduced for the 1+log trumpet compared to the 
maximal trumpet, although on the given scale there is not much
difference between the two types of trumpet data.

\subsection{Junk radiation and the gravitational wave signal}

\begin{figure}[tpb]
\centering
\includegraphics[width=0.45\textwidth]{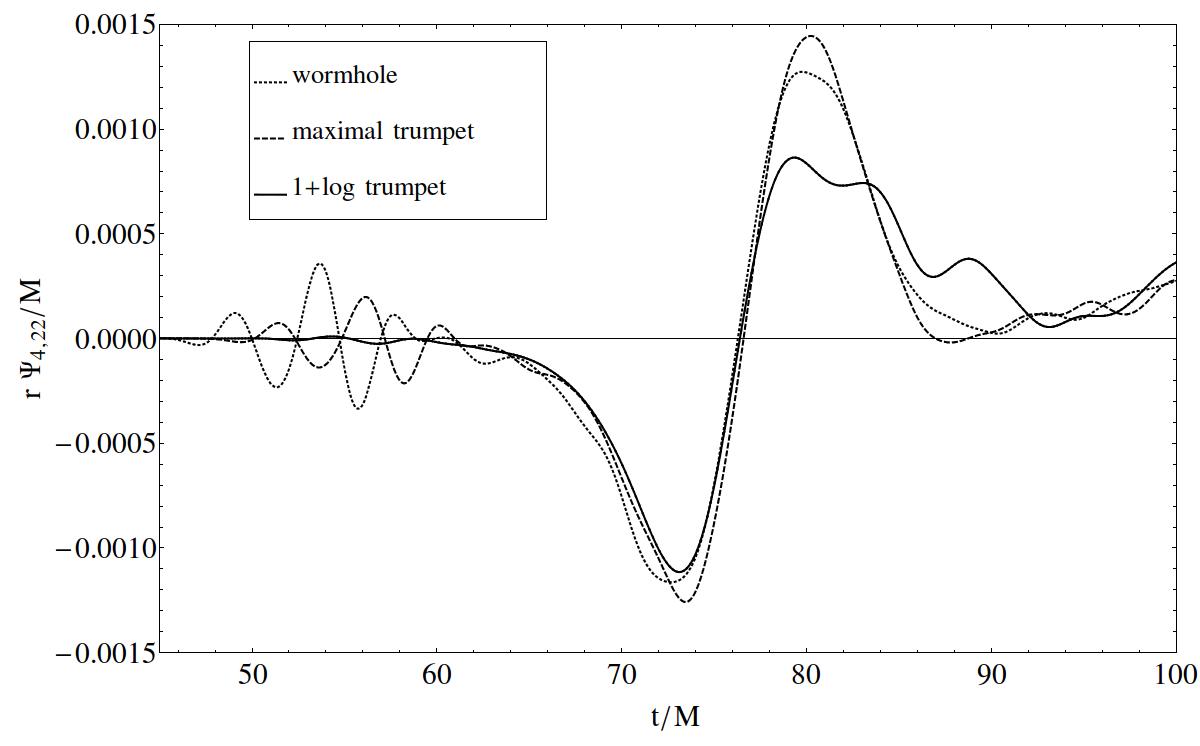}
\includegraphics[width=0.45\textwidth]{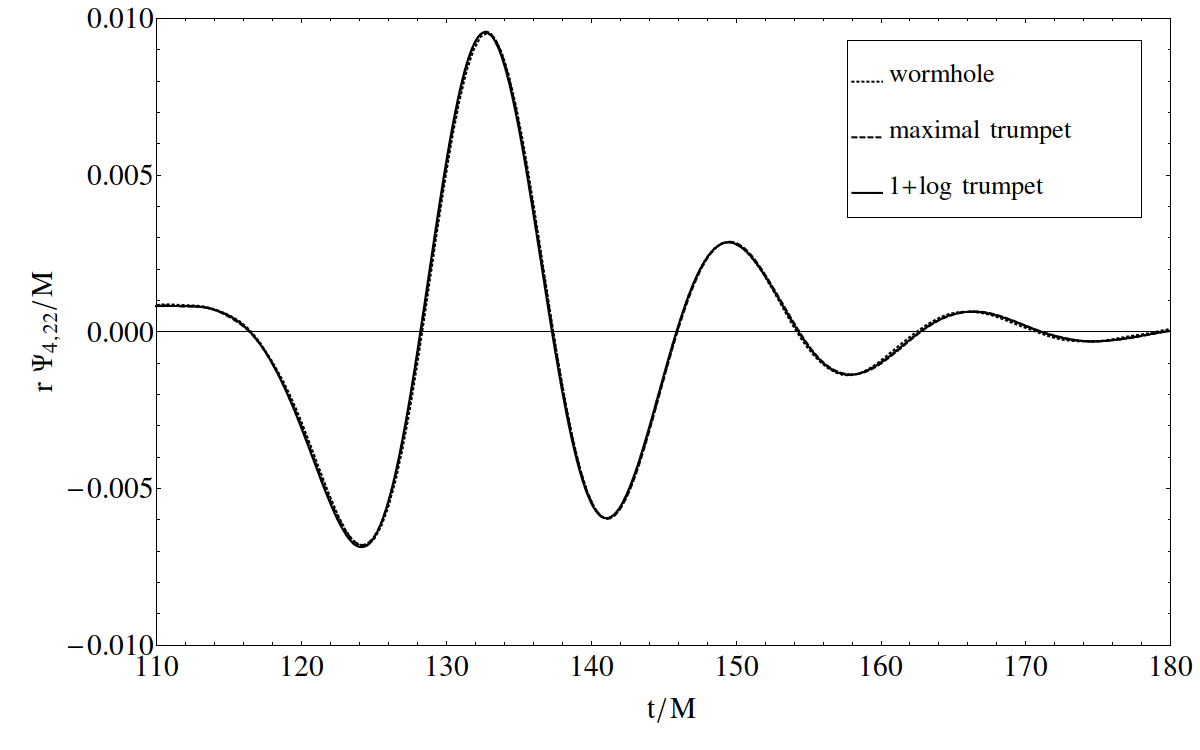}
\caption{Head-on collision of two black holes.
Shown is the junk radiation 
(upper panel) and the gravitational wave signal at later times 
(lower panel). 
The time coordinate is shifted in such a way 
that characteristic points coincide.
}
\label{Figjunk}
\end{figure}

Since we are using conformally flat initial data, we do not avoid 
the production of junk radiation. 
This issue was discussed for maximal trumpets in \cite{HanHusOMu09},
and analogous arguments hold also for 1+log trumpets. 
Fig.~\ref{Figjunk} shows the gravitational radiation from the head-on 
collision of two punctures starting at $d=20m\approx10M$ in terms of 
the spin-weight $-2$, $l=2$, $m=2$ mode of $r \Psi_4$
computed at an extraction radius of $r_{ex}=75M$. 
The amount of junk radiation (upper panel) for wormholes and trumpets
is approximately the same.  
The figure suggests that some features of the early
wave pulse are due to conformal flatness, while others (the leading oscillations
for wormhole data that are absent from 1+log trumpet data) may be residues
of the early gauge dynamics. Fig.~\ref{Figjunk} can be compared with 
Fig.~12 of \cite{HanHusOMu09}, which shows a similarly small difference 
in the junk radiation between a maximal trumpet and a wormhole, including
small oscillations for wormhole data.
A significant reduction of junk radiation can be achieved by computing
non-conformally-flat initial data, e.g.\ 
\cite{TicBruCam02,YunTicOwe05,LovOwePfe08,Lov08,KelTicZlo10,ReiTic12}.

Another important question is if the different types of initial data
describe the same physical system.  For this purpose we have a
closer look at the gravitational wave signal at later times in
Fig.~\ref{Figjunk} (lower panel). The physically relevant part of the
gravitational wave signal is nearly identical on this scale.

There are two obvious reasons why the signals can not be identical.
Firstly, we have not solved the momentum constraint, and therefore we
use a setup which deviates slightly from an exact solution of
Einstein's field equations.  Secondly, using the same bare initial
masses for wormholes, maximal trumpets, and 1+log trumpets leads to
different ADM-masses. We can rescale by the ADM mass $M$, but this
leads to small differences in the rescaled initial separation.
However, the results agree quite well even without fine tuning.

\section{Conclusion}
\label{Con}

The aim of this paper was to solve the Hamiltonian constraint for
1+log trumpets, as opposed to maximal trumpets or wormholes. Since
1+log trumpet initial data is constructed in the approximately stationary, 
quasi-equilibrium gauge of evolutions using the standard moving puncture method, such
initial data is a natural choice that can be expected to minimize the
initial gauge dynamics.

As a general strategy to address a pole singularity in the conformal
factor, we suggest considering negative powers of the conventional
conformal factor, see also \cite{Gun10a,Bau12a,Die12}. In fact, the original
additive puncture method fails for 1+log trumpet punctures, while we showed
that regularizing the conformal factor by using its inverse succeeds
both for wormholes and trumpets. Note that the character of the Hamiltonian
constraint equation is different for wormholes, maximal trumpets, and
1+log trumpets.  The novelty of the present work is a working scheme
for the superposition of two 1+log Schwarzschild trumpets, which is
the first treatment of the $K\neq0$ case in this context.

One open issue is the ad hoc approach to the momentum constraint, for
which various approximate solutions were constructed. It is encouraging
that in actual evolutions the violation of the momentum constraint reached
comparable levels even for constraint-solved wormhole data. The evolutions
also indicated that the 1+log trumpet data indeed contain fewer artifacts.
It remains to be investigated whether similar methods can be applied to the 
solution of the momentum constraint.

With the basic superposition of two 1+log Schwarzschild trumpets
available, a followup project is the inclusion of momentum and
spin. This can in principle be achieved by adding Bowen-York extrinsic
curvature to the (non-vanishing) extrinsic curvature of the head-on
1+log trumpet configuration. However, in order to avoid the artificial
radiation in Bowen-York data, we would prefer a method that is based
on the quasi-equilibrium state of orbiting trumpets in the moving
puncture gauge.

Part of this exercise is academic because wormhole punctures quickly
and reliably evolve into quasi-stationary 1+log punctures. There are
advantages if the early part of the waveform is important, but this
may not be essential in practical terms. Still, from a theoretical
point of view, constructing moving puncture initial data is worthwhile
since it would resolve the puzzling state of affairs that one of
the most successful slicings of binary black spacetimes can currently
only be found by performing actual evolutions, without having an
approximate initial data construction available. \\

\begin{acknowledgments}

It is a pleasure to thank M. Ansorg, D. Hilditch, and 
J. Gundermann for helpful discussions. We are grateful to T. Baumgarte and N. K. 
Johnson-McDaniel for comments on the manuscript. 
This work was supported in part by DFG grant SFB/Transregio~7
``Gravitational Wave Astronomy''. 
T.D. was supported by the
Graduierten-Akademie Jena and the Studienstiftung des deutschen
Volkes. Computations were performed at the Quadler cluster of the
Institute of Theoretical Physics of the University of Jena
and on SuperMUC of the Leibniz Rechenzentrum. 

\end{acknowledgments}

\end{document}